\documentclass[12pt]{iopart}

\usepackage{graphicx} 

\begin{document}

\title[Complex phases in weak measurements]{On the role of complex phases in the quantum statistics of weak measurements}

\author{Holger F Hofmann}

\address{Graduate School of Advanced Sciences of Matter, Hiroshima University,
Kagamiyama 1-3-1, Higashi Hiroshima 739-8530, Japan} 
\address{JST, CREST, Sanbancho 5, Chiyoda-ku, Tokyo 102-0075, Japan}
\ead{hofmann@hiroshima-u.ac.jp}

\begin{abstract}
Weak measurements performed between quantum state preparation and post-selection result in complex values for self-adjoint operators, corresponding to complex conditional probabilities for the projections on specific eigenstates. In this paper, it is shown that the complex phases of these weak conditional probabilities describe the dynamic response of the system to unitary transformations. Quantum mechanics thus unifies the statistical overlap of different states with the dynamical structure of transformations between these states. Specifically, it is possible to identify the phase of weak conditional probabilities directly with the action of a unitary transform that maximizes the overlap of initial and final states. This action provides a quantitative measure of how much quantum correlations can diverge from the deterministic relations between physical properties expected from classical physics or hidden variable theories. In terms of quantum information, the phases of weak conditional probabilities thus represent the logical tension between sets of three quantum states that is at the heart of quantum paradoxes. 
\end{abstract}


\maketitle

\section{Introduction}

Weak measurements are a method to determine the statistical properties of quantum systems between state preparation and a specific post-selected measurement result \cite{Aha88}. Originally, weak measurements met with a considerable amount of skepticism due to their seemingly paradoxical and unconventional nature \cite{Duc89,Leg89}. However, there has recently been a renewed interest in weak measurements, not only motivated by new experimental possibilities \cite{Pry05,Mir07,Iin11}, but also because weak measurements may clarify fundamental issues in quantum mechanics by indicating possible experimental resolutions of quantum paradoxes \cite{Mir07,Tol07,Tol10,Lun10,Res04,Aha02,Lun09,Yok09,Jor06,Wil08,Gog11}. Such resolutions of quantum paradoxes are usually based on the interpretation of weak measurement statistics in terms of negative probabilities whose averages can reproduce the experimentally observed violation of inequalities \cite{Hos10,Hof09,Hof10}. A more detailed analysis of the strange features of weak measurements may therefore lead to a better understanding of the essential differences between quantum and classical statistics.

As early as 1995, Steinberg pointed out that weak measurements provide a natural definition of conditional probabilities in quantum mechanics \cite{Ste95}. However, the mathematically consistent definition of such weak conditional probabilities results in complex numbers originating from the quantum coherences of the initial and final states. In terms of a purely statistical interpretation, the complex phases of weak conditional probabilities seem to pose a problem, since it is not immediately obvious how complex values contribute to any experimentally observable statistics. In this paper, I therefore take a closer look at the actual physics described by the complex conditional probabilities obtained in weak measurements. The results show that the complex phases of weak conditional probabilities describe the responses of the transition probability from initial to final state to unitary transformations commuting with the intermediate measurement. This makes it possible to interpret complex conditional probabilities in terms of transformation dynamics. In particular, the complex phase obtained for a specific intermediate measurement result $m$ defines the action $S(m)$ of a unitary transformation that maximizes the overlap between the initial and the final state. The change of $S(m)$ with $m$ describes a physical distance between the initial and the final state around a specific result $m$. Specifically, fast oscillations of the complex conditional probabilities associated with a rapidly changing action $S(m)$ indicate a significant separation between initial and final state within the respective range of $m$ values. Thus, the complex phases of weak conditional measurements describe situations where the initial and final state appear to attribute different physical properties to the system at the intermediate state $m$, suggesting a classical contradiction between the physics defined by the three states. Effectively, quantum mechanics seems to replace the conditional probability of zero for logical contradictions between sets of three statements with complex probabilities that average out when observed with sufficiently low resolution. 

In general, the complex phases of weak conditional probabilities convey specific information on the transformation dynamics of the output statistics by softening the logical relations between quantum states. The specific non-classical correlation between conditional statistics and transformation dynamics is expressed by non-zero complex probabilities for classically inconsistent sets of states, where the complex phase is given by the action of the transformation that minimizes the logical contradictions between the three states. In the context of quantum information and quantum paradoxes, the complex phases of weak conditional probabilities can therefore be understood as a measure of the logical tension between sets of three non-orthogonal quantum states. While classical statistical expectations are reproduced when the logical tension is low, logical tensions larger than $\pi/2$ result in the negative conditional probabilities that can be used to characterize quantum paradoxes \cite{Res04,Aha02,Lun09,Yok09,Jor06,Wil08,Gog11}. The unification of transformation dynamics and conditional probabilities in terms of the logical tension between three states may thus lead to a better understanding of the non-classical properties of quantum statistics.

The rest of the paper is organized as follows. In section \ref{sec:cwv}, it is pointed out that the real part of weak values are determined by weak measurements, while the imaginary part is determined by weak unitary transformations. In section \ref{sec:unitary}, it is shown how the response to arbitrarily strong unitary transformations can be predicted from complex weak conditional probabilities. In section \ref{sec:max}, the complex phase of weak conditional probabilities is identified with the action of a transformation that minimizes the differences between initial and final state at all intermediate results $m$. In section \ref{sec:phasespace}, the $m$-dependent distance between initial and final state is illustrated for the case of particle position. It is shown that the derivative of the action $S(x)$ corresponds to the momentum difference between initial and final state at $x$. In section \ref{sec:logtense}, the implications of the transformation distance given by the complex phase for the logical relation between three quantum states is considered and the concept of logical tension is introduced. In section \ref{sec:classlimit}, the transition to the classical limit of mixed states is analyzed and inequalities for the predictions of transformation statistics from complex conditional probabilities are derived. Finally, the conclusions are summarized in section \ref{sec:concl}. 

\section{Complex weak values}
\label{sec:cwv}

Weak values can be obtained when the measurement interaction is so low that the back-action effects of the measurement can be neglected. As discussed in \cite{Hof10}, weak measurements can be efficiently represented by measurement operators for the actual outcomes $\mu$ of the weak measurement,
\begin{equation}
\label{eq:measurement}
\hat{E}_\mu = \sqrt{w_\mu}\left(1+\epsilon_\mu \hat{A}\right),
\end{equation}
where $w_\mu$ gives the probability distribution of the meter readout $\mu$ before the weak interaction and $\epsilon_\mu$ describes the weak coupling that results in small modifications of the meter statistics based on the value of the observable $\hat{A}$. 
The measurement is weak if $\epsilon_\mu$ is so small that quadratic terms can always be neglected. When applied to an initial state $\mid i \rangle$, the measurement operator $\hat{E}_\mu$ modifies the statistical weight of the eigenstates of the observable $\hat{A}$ according to their eigenvalues. As a result, the output probability $p(\mu)$ is modified in proportion to the average value of $\hat{A}$. For the initial state $\mid i \rangle$, this modification is given by
\begin{equation}
\label{eq:average}
p(\mu|i)=\langle i \mid \hat{E}_\mu^2 \mid i \rangle = w_\mu \left(1+2 \epsilon_\mu \langle i \mid \hat{A} \mid i \rangle\right).
\end{equation}
For sufficiently small couplings $\epsilon_\mu$, the measurement does not change the quantum statistics of a final measurement. It is therefore possible to treat the post-selection of a final result $\mid f \rangle$ as a condition that is completely independent of the measurement dynamics (see \cite{Hof10} for details). According to Bayesian statistics, the conditional probability for the weak measurement outcome $\mu$ is then
\begin{equation}
\label{eq:weakaverage}
p(\mu|if) = \frac{|\langle f \mid \hat{E}_\mu \mid i \rangle|^2}{|\langle f \mid i \rangle|^2}
\approx w_{\mu} \left(1+2 \epsilon_{\mu} \mathrm{Re}\left(\frac{\langle f \mid \hat{A} \mid i \rangle}{\langle f \mid i \rangle} 
\right)\right),
\end{equation}
where the quadratic terms in $\epsilon_\mu$ have been neglected. As the comparison of Eq.(\ref{eq:average}) and Eq.(\ref{eq:weakaverage}) shows, the conditional average of $\hat{A}$ is given by the real part of the weak value,
\begin{equation}
\langle \hat{A} \rangle_{\mathrm{weak}} = \frac{\langle f \mid \hat{A} \mid i \rangle}{\langle f \mid i \rangle}.
\end{equation}
Significantly, the measurement probability given by Eq.(\ref{eq:weakaverage}) is completely symmetric in time, so that the imaginary part of the weak value does not contribute to weak measurements based on self-adjoint measurement operators.

It has been pointed out that the imaginary part of the weak value can be observed in system-meter interactions if the imaginary part is identified with a ``shift'' in the momentum of the pointer \cite{Jos07}. However, the pointer momentum is a conserved quantity in the system-meter interaction, so the explanation of the change in output probabilities in terms of a dynamic change of momentum is a misinterpretation. As pointed out in \cite{Hof11}, the (unchanged) momentum of the pointer represents the measurement back-action associated with the force that the pointer exerts on the system. 
In the quantum formalism, this force is represented by a parameter $\phi$ in the unitary transformation that expresses the transformation of the quantum system caused by the action of the force. 
Although the weak back-action effects average out when the fluctuating momentum is unknown, there is a correlation between the fluctuations of the force $\phi$ and the fluctuations of the final measurement result $f$ due to the dynamic response of the system to weak unitary transformations generated by the observable $\hat{A}$. 
The unitary operator that describes the statistical effects associated with sufficiently small forces $\phi$ has the form
\begin{equation}
\label{eq:unitary}
\hat{U}_\phi = \exp(-i \phi \hat{A}) \approx (1-i \phi \hat{A}).
\end{equation}
Comparison of Eq.(\ref{eq:measurement}) and Eq.(\ref{eq:unitary}) shows why the response to a weak unitary appears to be the imaginary part of the weak measurement result. However, it should not be forgotten that the physics of unitary transformations is quite different from the physics of measurement. In fact, classical physics clearly distinguishes the two concepts: a transformation changes the physical properties without any change to the available information, while a measurement changes the available information, ideally without changing any physical properties. The problem is that quantum mechanics makes ideal measurements impossible. However, weak measurements come close to the ideal case, so it is possible to identify the real parts of weak values with back-action free measurement statistics, and the imaginary parts with information free transformations. Specifically, the imaginary part of the weak value is equal to half the logarithmic derivative of the final probability $p(f|i)$ in $\phi$ \cite{Hof11},
\begin{equation}
\label{eq:logderiv}
\frac{1}{2}\frac{\partial}{\partial \phi} \ln \left(p(f|i)\right) = 
 \mathrm{Im}\left(\frac{\langle f \mid \hat{A} \mid i \rangle}{\langle f \mid i \rangle} 
\right).
\end{equation}
Thus, the imaginary part of the weak value is a measure of the differential response of the final measurement result $f$ to small phase shifts generated by $\hat{A}$. In this sense, the complex phase of weak values relates the statistical averages given by the real part to the conjugate dynamic responses given by the imaginary part. In the following, I will show that this relation between statistics and dynamics provides the key to a deeper understanding of weak measurement statistics.

\section{Weak conditional probabilities and unitary transformations}
\label{sec:unitary}

Initially, weak values attracted attention because they can lie outside the spectrum of eigenvalues observed in strong measurements. However, it is always possible to represent an operator by its spectral decomposition. The problem of unusual weak values can then be reduced to the perhaps more fundamental problem of unusual statistics. Specifically, the weak value of an observable $\hat{A}$ with eigenvalues $A_m$ and eigenstates $\mid m \rangle$ is
\begin{equation}
\label{eq:specstat}
\langle \hat{A} \rangle_{\mathrm{weak}} =
\sum_m A_m \frac{\langle f \mid m \rangle \langle m \mid i \rangle}{\langle f \mid i \rangle}.
\end{equation}
This weak value can be interpreted as an average defined by weak conditional probabilities of
\begin{equation}
\label{eq:weakcond}
p(m|if)=\frac{\langle f \mid m \rangle \langle m \mid i \rangle}{\langle f \mid i \rangle}.
\end{equation}
Weak conditional probabilities explain the weak values of any operator with eigenstates $\mid m \rangle$. Thus, weak conditional probabilities provide a consistent description of the non-classical statistics observed in weak measurements \cite{Hos10,Hof10,Ste95}. In particular, weak conditional probabilities provide an empirical tool for the investigation of non-classical correlations between measurement results that cannot be obtained jointly. As shown in a number of recent experiments \cite{Mir07,Tol07,Tol10,Lun10,Res04,Aha02,Lun09,Yok09,Jor06,Wil08,Gog11}, it is then possible to explain quantum paradoxes in terms of negative conditional probabilities for the weakly measured alternatives $m$. In such demonstrations of non-classical statistics, weak conditional probabilities establish a link between the conventional representation of quantum coherence as a wavelike property and classical probability theory. Eq.(\ref{eq:weakcond}) expresses this fundamental relation between the complex conditional probabilities obtained in weak measurements and the quantum coherence of Hilbert space. In particular, the post-selection of a final state $\mid f \rangle $ which is an equal superposition of all intermediate results $\mid m \rangle$ results in conditional probabilities $p(m|if)$ that are directly proportional to the complex amplitudes $\langle m \mid i \rangle$ of the initial state. As demonstrated in a recent experiment \cite{Lun11}, this proportionality can be used to realize a direct measurement of the wavefunction of a quantum state. 

Since the wavefunction is the more familiar theoretical concept, the discussion in \cite{Lun11} ignores the connection between this interpretation of the wavefunction as a specific conditional probability and the negative conditional probabilities observed in quantum paradoxes. However, the significance of quantum paradoxes as indicators of the non-classical properties of quantum information implies that the appearance of negative probabilities in the conditional statistics of quantum measurement might be the most significant feature of quantum coherence. In addition, the conditional probabilities determined in weak measurements have an imaginary part that is related to the statistical response to weak unitaries give by Eq.(\ref{eq:logderiv}). Quantum coherence thus describes a combination of static probabilities with elements of unitary transformation dynamics that has no analogy in classical statistics. Consequently, it should be possible to identify the fundamental difference between quantum statistics and classical statistics by analyzing the role of the complex phase in weak conditional probabilities. 
 
The following discussion shows how quantum coherence can be interpreted in a statistical context - and how an image of the complex wavefunction of a quantum system can appear in the quantum mechanical limit of conditional probabilities when the post-selected state is a (not necessarily equal) superposition of the alternative outcomes investigated in the weak measurement. The result presented in the following can therefore clarify and generalize the physical principles behind the direct observation of quantum coherence reported in \cite{Lun11} and help to identify the relation with quantum paradoxes.

Eq. (\ref{eq:specstat}) indicates that the imaginary part of a weak value can be expressed in terms of a weighted sum over the imaginary parts of weak conditional probabilities. This means that weak conditional probabilities have to be complex because the imaginary part is required to describe all differential changes to the final measurement statistics caused by unitary transformations with eigenstates $\mid m \rangle$. Specifically, the logarithmic derivative of the final probability $p(f|i)$ in $\phi$ can now be expressed as
\begin{equation}
\label{eq:diff}
\frac{1}{2}\frac{\partial}{\partial \phi} \ln \left(p(f|i)\right) = \sum_m A_m
\mathrm{Im}\left(p(m|if)\right).
\end{equation}
The imaginary parts of weak conditional probabilities thus provide a detailed description of the linear response to unitaries that commute with (and hence conserve) the projectors $\mid m \rangle \langle m \mid$. 
However, unitary dynamics are not limited to the differential changes in $\phi$ that define imaginary weak values. For arbitrarily large parameters $\phi$, the unitary operation can be represented by the spectral decomposition that assigns an action of $\phi A_m$ to each eigenstate projector $\mid m \rangle\langle m \mid$. By identifying the terms of this spectral decomposition with weak conditional probabilities, it is possible to derive the general relation between complex probabilities and the effects of unitary transformations. Specifically, the effect of a unitary operation $\hat{U}_\phi=\exp(-i \phi \hat{A})$ on the output probabilities $p(f|i)$ for an initial state $\mid i\rangle$ can be written as
\begin{equation}
\label{eq:unitrans}
|\langle f \mid \hat{U}_\phi \mid i \rangle|^2 = \left|
\sum_m \exp(-i \phi A_m)
\frac{\langle f \mid m \rangle \langle m \mid i \rangle}{\langle f \mid i \rangle}
\right|^2 |\langle f \mid i \rangle|^2.
\end{equation}
Using weak measurement statistics, it is now possible to interpret this transformation in terms of complex conditional probabilities. Specifically, the output probability $p(f;\phi)=|\langle f~\mid \hat{U}_\phi \mid~i~\rangle|^2$ can be obtained from the complex conditional probabilities $p(m|if)$ obtained near $\phi=0$ using the relation 
\begin{equation}
\label{eq:transcorr}
p(f;\phi) = \left|\sum_m \exp(-i \phi A_m)
p(m|if)
\right|^2 p(f;0).
\end{equation}
The dependence of the output probability $p(f;\phi)$ on the parameter $\phi$ can therefore be determined completely using only the weak effects observed around $\phi=0$.

In general, a transformation generated by $\hat{A}$ conserves the value of $m$ but changes the value of $f$. Experimentally, the conditional probability $p(m|if)$ is obtained by post-selecting only systems with a specific value of $f$. After the unitary transformation is applied, one would expect that some of the contributions to $p(f;\phi)$ originate from systems with different values of $f$. For example, the unitary transform could change the state $\mid f \rangle$ to an orthogonal state, so that $\hat{U}_\phi^\dagger\mid f \rangle$ represents an experimentally distinguishable alternative outcome $\mid g \rangle$. In this case, the probability of $g$ can be derived from the weak conditional probabilities of $f$ using Eq.(\ref{eq:transcorr}). In classical statistics, there would be no reason to assume that the conditional probabilities at $f$ should be fundamentally related to the probability of obtaining a different measurement outcome $g \neq f$. Thus, the complex phases of weak conditional probabilities express a non-classical aspect of quantum statistics that has no obvious analogy in classical statistics.



\section{Maximizing the overlap of initial and final states}
\label{sec:max}

In classical physics, the transformation dynamics generated by $\hat{A}$ correspond to phase space trajectories that shift the phase space point by a distance of $\phi$ along a phase space contour $m$ with a constant value of $A_m$ for $\hat{A}$. In this analogy, quantum states correspond to classical phase space contours. The probability $p(f)$ for an initial state $i$ then originates from the intersection of two phase space contours. A transformation generated by $\hat{A}$ can modify the statistical overlap by reducing or increasing the distance between $i$ and $f$ along the different phase space contours $m$. Based on this analogy, it is possible to interpret the complex phases of weak conditional probabilities as an indication of the distance between the initial and the final state for transformations along the intermediate states $m$. 

In the Hilbert space formalism, the transformation is described by phase changes of $\phi A_m$ that correspond to the classical action of the transformation at $m$. 
Using Eq.(\ref{eq:transcorr}), it is easy to see that the maximal output probability is obtained when the action of $\phi A_m$ introduced by the transformation compensates the intrinsic phase of the weak conditional probability, so that the sum runs over the absolute values. In this case, the probability of finding the final outcome $\mid f \rangle$ is 
\begin{equation}
p(f;\mathrm{max.}) = \left(\sum_m |p(m|if)|
\right)^2 p(f;0).
\end{equation}
The unitary transform that achieves this maximal overlap between initial and final state while conserving $m$ can be defined in terms of an $m$-dependent action $S_m$,
\begin{equation}
\hat{U}_{\mathrm{max.}} = \sum_m \exp(-i S_m) \mid m \rangle \langle m \mid,
\end{equation} 
where the action $S_m$ is given by the complex phase of the weak conditional probability,
\begin{equation}
\label{eq:action}
S_m = \mathrm{Arg}\left(\frac{\langle f \mid m \rangle \langle m \mid i \rangle}{
\langle f \mid i \rangle}
\right) = \mathrm{Arg}\left( p(m|if) \right).
\end{equation}
The classical analogy suggests that the unitary transformation defined by the action $S_m$ moves the phase space point defined by the intersection of $i$ and $m$ along $m$ until it reaches the intersection of $m$ and $f$, where the distance between the two points is given by the gradient of the action in $m$. Thus, the complex phases of weak conditional probabilities actually seems to define a discrepancy between the physical properties described by the pairs of quantum states $(\mid i \rangle,\mid m \rangle)$ and $(\mid m \rangle,\mid f \rangle)$. 

Since complex probabilities have no classical analogy, it is interesting to find that they are related to the classical phase space structure that defines transformations of the system. Quantum mechanics appears to unify these two aspects of physics into a single formalism, where the assignment of phase space points must be replaced by complex conditional probabilities \cite{Hof11b}. As a result, the statistical relations between measurements that cannot be performed jointly may be paradoxical due to the negative conditional probabilities corresponding to the transformation dynamics that define the new relation between the physical properties. At the same time, it becomes possible to predict the effects of arbitrary transformations from the conditional statistics of a single measurement outcome $f$. It may therefore be possible to understand the non-classical features of quantum information in terms of the action of unitary transformations.


\section{Phase space illustration for continuous variables}
\label{sec:phasespace}

As the example of the time evolution generated by a Hamiltonian shows, the relation between the quantum mechanical action $S_m$ of an eigenstate component $\mid m \rangle$ in a unitary transformation corresponds to the classical action of that transformation in units of $\hbar$. In the classical case, the unitary transform can then be described in terms of a shift in the conjugate observable that parameterizes the classical phase space. For each value of $m$, the magnitude of this shift is then given by the $m$-derivative of $S_m$. In quantum mechanics, the values of $m$ are usually discrete and the definition of conjugate observables is difficult. On the other hand, continuous variables such as position and momentum  preserve much more of the classical phase space structure. It may therefore be useful to take a look at the weak conditional probabilities of position for a more intuitive picture of the differences between initial and final states described by their complex phases.

If the weak conditional probability density of position is given by $p(x|if)$,
then the transformation $\hat{U}_{\mathrm{max.}}$ that maximizes the overlap of initial and final state is 
\begin{equation}
\hat{U}_{\mathrm{max.}} = \int \exp\left(-i \; \mathrm{Arg}\left( \frac{\langle f \mid x \rangle \langle x \mid i \rangle}{
\langle f \mid i \rangle}\right)\right)
\mid x \rangle \langle x \mid. 
\end{equation}
In terms of Hamiltonian dynamics, this transformation corresponds to the 
application of a potential $V(x)$ over a time $t$, such that $V(x) t = \hbar S(x)$. The classical change of momentum caused by this transformation at $x$ would be 
\begin{equation}
\Delta P(x) = - \hbar \frac{\partial}{\partial x} S(x)
\end{equation}
In quantum mechanics, it is possible to express the $x$-derivative on the right hand side of the equation in terms of the momentum operator $\hat{P}$. Interestingly, this results in an identification of the $x$-derivatives of phase with the real parts of the weak values of momentum at $x$. Specifically,
\begin{eqnarray}
- \hbar \frac{\partial}{\partial x} S(x) &=& \hbar \frac{\partial}{\partial x} \mathrm{Arg}\left( \langle x \mid f \rangle \right) - \hbar \frac{\partial}{\partial x} \mathrm{Arg}\left( \langle x \mid i \rangle \right)
\nonumber \\
&=& \mathrm{Re}\left(\frac{\langle x \mid \hat{P} \mid f \rangle}{\langle x \mid f \rangle} \right) - \mathrm{Re}\left(\frac{\langle x \mid \hat{P} \mid i \rangle}{\langle x \mid i \rangle} \right).
\end{eqnarray}
Thus continuous variable quantum mechanics confirms the intuitive notion that the transformation $\hat{U}_{\mathrm{max.}}$ maximizes the overlap of initial and final state by minimizing the difference in the conjugate observable $\hat{P}$ at position $x$. 

Since the momentum difference $\Delta P(x)$ is equal to the $x$-derivative of the action $\hbar S(x)$, it is tempting to identify the action directly with the phase space integral between $\mid i \rangle$, $\mid f \rangle$ and $\mid x \rangle$, where the phase space representations of $\mid i \rangle$ and $\mid f \rangle$ are given in terms of their weak value momenta at $x$. However, some care should be taken since the integration results in a constant that needs to be defined by the normalization of the complex conditional probabilities to $1$. A specific example may help to illustrate the point. Consider a free particle of mass $m$. Its position at various times can be expressed in terms of position $\hat{x}$ and momentum $\hat{P}$ at time $t=0$. If $\mid i \rangle$ and $\mid f \rangle$ are eigenstates of particle position $x=0$ at times $t=-\tau/2$ and $t=\tau/2$, they are defined by the operator relations 
\begin{eqnarray}
\left(\hat{x}-\frac{\tau}{2 m} \hat{P}\right) \mid i \rangle &=& 0 
\nonumber \\
\left(\hat{x}+\frac{\tau}{2 m} \hat{P}\right) \mid f \rangle &=& 0.
\end{eqnarray}
Note that each point in the phase space defined by $\hat{x}$ and $\hat{P}$ refers to a complete trajectory in time. Thus, the time evolution of the system is fully accounted for by the definition of the initial and final states at $t=0$ and explicit descriptions of time dependences are unnecessary. It is easy to see that the weak values of momentum conditioned by a measurement of position $\hat{x}$ at $t=0$ are $P_i(x)=2 m x/\tau$ and $P_f(x)= - 2 m x/\tau$. In phase space, the initial and the final states are therefore represented by straight lines with opposite slopes intersecting at $(x=0;P=0)$. 

Using an arbitrary normalization length of $L$, the wave functions of the initial and final states can be given as 
\begin{eqnarray}
\langle x \mid i \rangle &=& \frac{1}{\sqrt{L}} \exp\left(i \frac{m}{\hbar \tau} x^2 \right)
\nonumber \\
\langle x \mid f \rangle &=& \frac{1}{\sqrt{L}} \exp\left(- i \frac{m}{\hbar \tau} x^2\right).
\end{eqnarray}
The weak conditional probability density determined from these two wave functions is
\begin{equation}
\label{eq:xif}
p(x|if) = \sqrt{\frac{2 m}{\pi \hbar \tau}} \exp \left(i \frac{2 m}{\hbar \tau} x^2 - i \frac{\pi}{4}\right).
\end{equation}
Interestingly, the complex phase $S(x)$ of this weak conditional probability is $-\pi/4$ for the classical solution at $x=0$. As a result of this phase shift, the range of $x$ values contributing positive real parts to the total probability is broadened to include $x$-values with actions up to $(3\pi /4)\hbar$ higher than the minimal action at $x=0$.

\begin{figure}[th]
\begin{picture}(440,200)
\put(0,0){\makebox(440,200){\vspace*{-3.5cm}
\scalebox{0.7}[0.7]{
\includegraphics{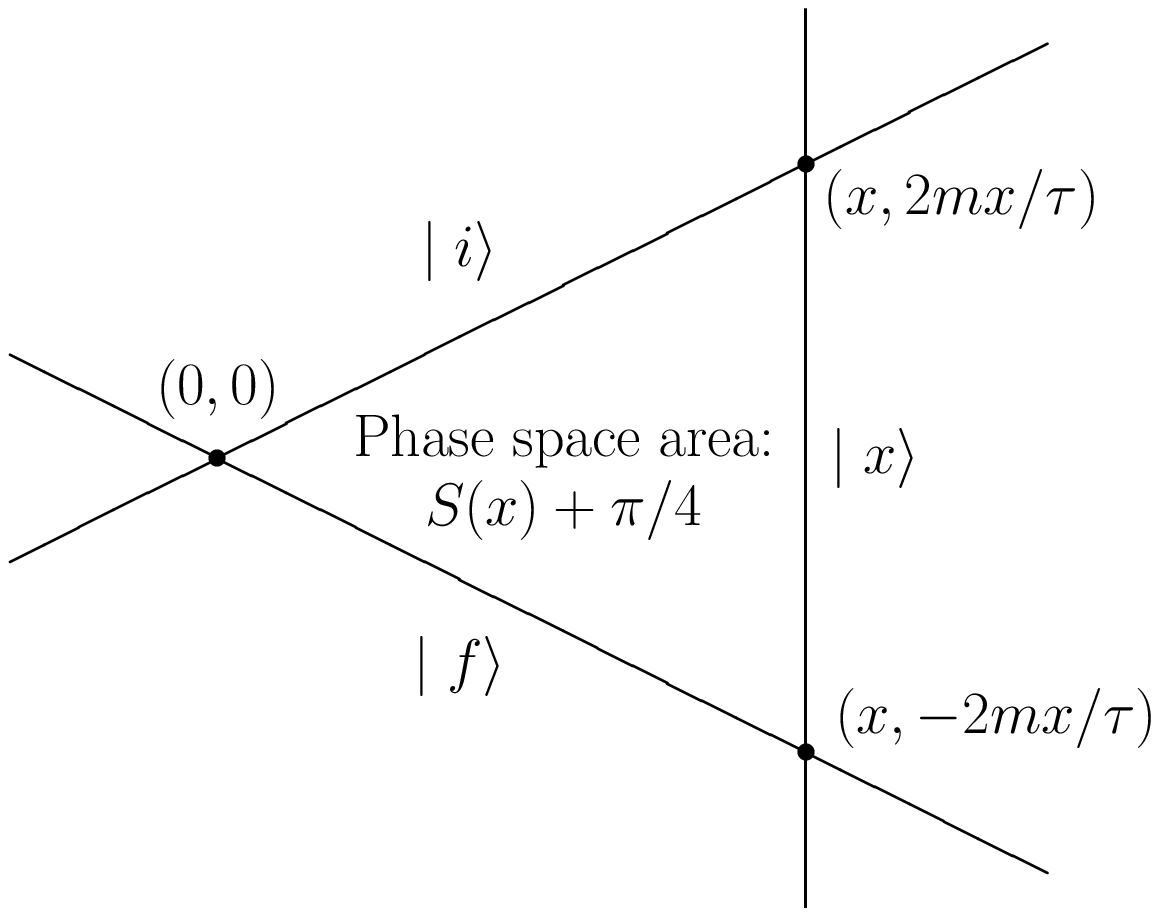}}}}
\end{picture}
\caption{\label{fig1} Phase space illustration of weak conditional probabilities for an initial eigenstate of $\hat{x}-\tau \hat{P}/2m$ and a final eigenstate of $\hat{x}+\tau \hat{P}/2m$. The phase space area enclosed by the three states is equal to the complex phase $S(x)$ of the weak conditional probability plus $\pi/4$.}
\end{figure}

The phase space geometry corresponding to the weak conditional probability in Eq.(\ref{eq:xif}) is illustrated in Fig. \ref{fig1}. Each quantum state corresponds to a straight line, and all three states form a triangle. Classically, the crossing point of initial and final state at $(x=0, P=0)$ would correspond to the trajectory defined by initial and final conditions, and the conditional probability for $x$ would be a delta function around $x=0$. Obviously, quantum mechanics relaxes this tight deterministic relation. Instead, the quantitative disagreement between the three different trajectories defined by the three pairs of quantum states results in a complex phase proportional to the total phase space area between the three non-identical phase space points, minus a normalization phase of $\pi/4$. In the macroscopic limit, classical determinism is recovered, because a low resolution measurement of $x$ will average over several periods of oscillation of the complex phase, leaving only the slowly varying positive contributions to the conditional probability around $x=0$. However, a resolution at the pure state level will always provide the complete information on the phase space distances between the quantum states concerned. 

\section{Logical tension between three quantum states}
\label{sec:logtense}

The phase space analysis shows that weak conditional probabilities are non-zero even though the initial, intermediate and final state do not intersect at a common phase space point. If there is a non-zero pairwise overlap of the three states, quantum mechanics expresses the conditional relation of all three states in terms of the complex phase. Classical determinism is replaced with phases close to zero, while classical contradictions are replaced with rapidly oscillating phases. In this sense, the complex phase associated with unitary transformation functions provides a quantitative measure of the non-classical logical relation between quantum statements. Specifically, it may be useful to consider the complex phase of weak conditional probabilities as the measure of logical tension between sets of three quantum states. If the logical tension is low, conditional and joint probabilities are positive and the rules of classical statistics apply. On the other hand, logical tensions above $\pi/2$ result in negative joint probabilities such as those observed in quantum paradoxes \cite{Res04,Aha02,Lun09,Yok09,Jor06,Wil08,Gog11}. 

In general, the logical tension is a symmetric function of three states, invariant under permutations of the sequence of states. It is therefore appropriate to express it as
\begin{equation}
\label{eq:LT}
S(i,m,f) = \mathrm{Arg}\left(\langle f \mid m \rangle \langle m \mid i \rangle \langle i \mid f \rangle \right).
\end{equation}
Mathematically, this is the geometric phase of a cyclic product of the mutual overlaps of three states, also known as a Pancharatnam phase \cite{Tam09}. As such, it describes a rather fundamental feature of Hilbert space algebra. Due to the analysis given in the previous sections, this geometric phase can now be identified with a mismatch in the physical properties defined by the three states. 

In the case of continuous variables, the logical tension between three quantum states is a function of the phase space area enclosed by the three states as shown in Fig. \ref{fig1}. However, quantum information is most commonly formulated in terms of two level qubit systems. For such systems, quantum states can be illustrated by points on the Bloch sphere. As explained in \cite{Tam09}, the complex phase defined by Eq.(\ref{eq:LT}) is then equal to half the area of the geodesic triangle defined by the three points on the sphere. Like before, the logical tension between three states is illustrated by a triangle, indicating the symmetry of the three states.

\begin{figure}[th]
\begin{picture}(440,210)
\put(0,0){\makebox(440,200){\vspace*{-3cm}
\scalebox{0.7}[0.7]{
\includegraphics{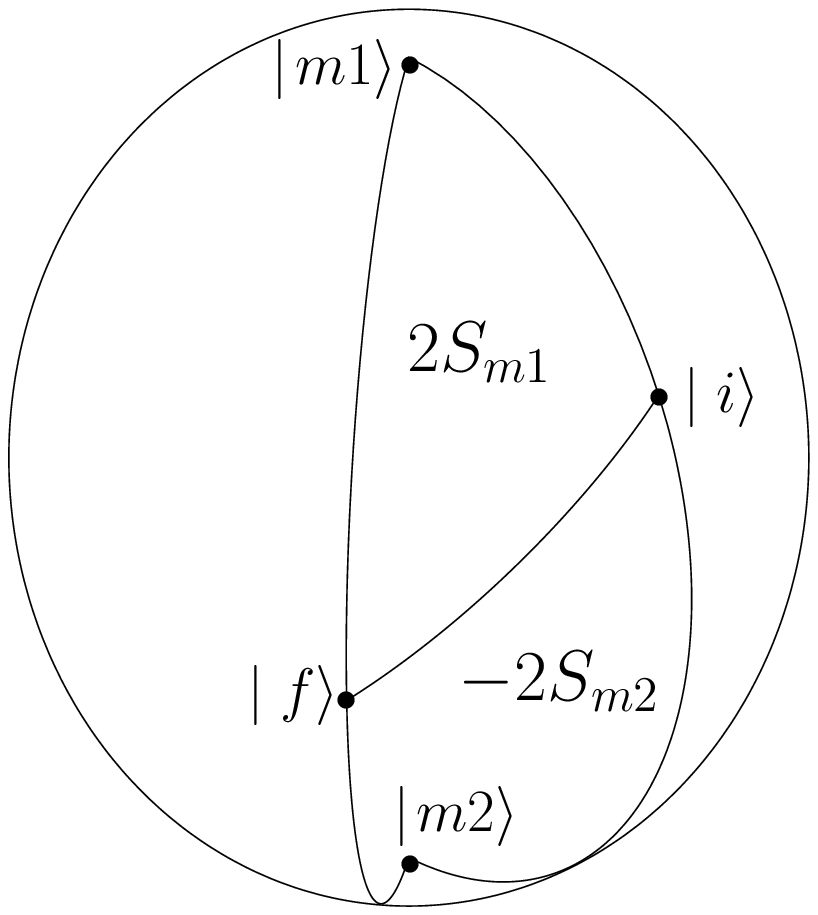}}}}
\end{picture}
\caption{\label{fig2} Logical tension on the Bloch sphere. For any combination of three states, the phase of the weak conditional probabilities is equal to half the area of the geodesic triangle defined by the three states. For a pair of orthogonal intermediate states, the phase difference is equal to half the area enclosed by the geodesics through $\mid i \rangle$ and through $\mid f \rangle$, corresponding to the angle of rotation between the two geodesics. Thus, rotating $\mid i \rangle$ into the same half plane as $\mid f \rangle$ relaxes the logical tension to zero.}
\end{figure}

The complete conditional probability for qubits is obtained by just two orthogonal results $m$ represented by opposite poles of the Bloch sphere as shown in Fig. \ref{fig2}. Due to the sequence of states, the logical tension has opposite sign for the two states. Therefore, the difference between the two logical tensions is equal to half the total area of the two triangles. In terms of the unitary operation $\hat{U}_{\mathrm{max.}}$, this phase difference describes a rotation around the $m$-axis that rotates $i$ into the same plane as $f$. Thus the logical tension is reduced to zero when all three states lie in the same half plane. Oppositely, logical tensions of $\pi$ are achieved when all three states lie in the same plane, but not in the same half plane. For the complete set of weak conditional probabilities, this means that the conditional probabilities are real and positive when $i$ and $f$ are on the same side of the $m$-axis, while one of the conditional probabilities is negative when $i$ and $f$ are on opposite sides. 

With regard to quantum paradoxes, classical statistical models work if the conditional probabilities are real and positive. Logical tension provides a natural quantitative expression for this condition. It seems therefore reasonable to define the classical limit as the limit of small logical tension. In the qubit case, this would suggest that initial and final states on the same side of the weak measurement axis could be considered classical. However, a simple rotation around the measurement axis will induce arbitrary amounts of logical tension. In general, Eq.(\ref{eq:transcorr}) indicates that all pure state systems can easily be transformed into states with high logical tensions between them. A more robust classical limit can only be obtained by considering mixed states, where the summation over probabilities with varying logical tensions may result in positive real values and small imaginary parts for all conditional probabilities obtained from unitary transformations in $m$. 

\section{Mixed states inequalities and classical limits}
\label{sec:classlimit}

As explained in section \ref{sec:unitary}, the precise prediction of output probabilities observed after arbitrarily large transformations of the system from weak conditional probabilities is a highly non-classical feature of quantum statistics. In the classical limit, this feature should be replaced with the expectation of statistical independence of the output probabilities obtained from strong transformations, limiting reliable predictions to the linear response to weak transformations given by the empirical definition of imaginary conditional probability in Eq.(\ref{eq:diff}). 

The proper classical limit is obtained when classical noise effects cover up the non-classical details of quantum statistics. Since Eq.(\ref{eq:weakaverage}) and Eq.(\ref{eq:logderiv}) provide an experimental recipe for obtaining the real and imaginary weak values, it is a straightforward matter to find the proper expressions of weak conditional probabilities in the presence of classical noise. For an initial state given by a density operator $\hat{\rho}_i$, the weak conditional probability is 
\begin{equation}
p(m|if)=\frac{\langle f \mid m \rangle \langle m \mid \hat{\rho}_i \mid f \rangle}{\langle f \mid \hat{\rho}_i \mid f \rangle}.
\end{equation}
This expression is mathematically equivalent to the pure state case if $\mid i \rangle$ is replaced with $\hat{\rho}_i \mid f \rangle$. 
As shown in \cite{Lun11}, the conditional probability for a well defined outcome $f$ can then be used to determine the wavefunction $\langle m \mid i \rangle$ of the initial state $\mid i \rangle$. However, this is not correct if the initial state is a mixed state. In this case, the method used in \cite{Lun11} will actually determine $\hat{\rho}_i \mid f \rangle$, an expression that describes only the fraction of the quantum state  $\hat{\rho}_i$ that is coherent with the final state $\mid f \rangle$. 
As the quantum coherence between $\mid f \rangle$ and other states decreases, $\hat{\rho}_i \mid f \rangle$ becomes more and more localized around $\mid f \rangle$, so that the conditional probability at $f$ becomes independent of the conditional probability at different final states $g$. 

For mixed states, the prediction of output probabilities for strong unitaries is complicated by the fact that the output probability cannot be separated into transitions between $i$ and $f$. Instead, the unitary $\hat{U}_\phi$ acts on both sides of the density operator $\hat{\rho}_i$. To obtain any predictions about the effects of the unitary, it is necessary to find a relation between this exact expression of the transformation dynamics and the expression $\langle f \mid \hat{U}_\phi \hat{\rho}_i \mid f \rangle$ that is related to the weak conditional probabilities. Such a relation can indeed be obtained in the form of a Cauchy-Schwarz inequality for the Hilbert space vectors $\hat{\rho}_i^{1/2} \hat{U}_\phi^\dagger \mid f \rangle$ and $\hat{\rho}_i^{1/2} \mid f \rangle$,
\begin{equation}
\langle f \mid \hat{U}_\phi \hat{\rho}_i \hat{U}_\phi^\dagger \mid f \rangle \geq \frac{\langle f \mid \hat{U}_\phi \hat{\rho}_i \mid f \rangle \langle f \mid \hat{\rho}_i \hat{U}_\phi^\dagger \mid f \rangle}{\langle f \mid \hat{\rho}_i \mid f \rangle}. 
\end{equation}
If the unitaries on the right hand side of the inequality are expressed by their spectral decompositions, the result is a sum over weak conditional probabilities that corresponds to the pure state case of Eq.(\ref{eq:transcorr}). Therefore, the mixed state inequality reads
\begin{equation}
\label{eq:tcorrlimit}
p(f;\phi) \geq \left|\sum_m \exp(-i \phi A_m)
p(m|if)
\right|^2 p(f;0),
\end{equation}
confirming that the pure state case given by Eq.(\ref{eq:transcorr}) is the ultimate limit of statistical predictions based on weak conditional probabilities. 

Eq.(\ref{eq:tcorrlimit}) shows that weak conditional probabilities always provide a correct prediction for the contributions of the quantum state $\hat{\rho}_i \mid f \rangle$ to the output statistics of the transformed system. However, other contributions will be added to the predicted ones, so that the predicted probabilities will only be a small fraction of the actual probabilities in the classical limit. To quantify the transition to the classical limit, it is useful to remember that the inequality above is always an equality for small $\phi$, since the changes of $p(f;\phi)$ that are linear in $\phi$ provide the very definition of imaginary weak values given by Eq. (\ref{eq:logderiv}). The effects of classical noise therefore first appear in the second derivative in the parameter $\phi$,
\begin{equation}
\frac{\partial^2}{\partial \phi^2} p(f;\phi\!=\!0) \geq - 2 \!\left(\!\sum_m A_m^2 \mathrm{Re}\left( p(m|if) \right) 
- \left|\sum_m A_m p(m|if)\right|^2 \right) p(f;0).
\end{equation}
Interestingly, the lower limit of this second derivative 
can be interpreted as 
minus two times the real part of the conditional uncertainty of $\hat{A}$, as defined by the complex values of the weak conditional probabilities. In the pure state limit, the equality of the two terms requires negative conditional uncertainties, since at least some second derivatives of $p(f;\phi=0)$ must be positive. In the classical limit, all uncertainties must be positive, so that positive second derivatives require a corresponding amount of additional (classical) noise. On the other hand, the magnitude of negative second derivatives is limited by the total uncertainty in $\hat{A}$, so the expectation for the classical limit is that the absolute ratio of the second derivative of the probability $p(f)$ in $\phi$ to the probability p(f) itself is everywhere larger than the uncertainty of $\hat{A}$ in the initial state. 
On the whole, these conditions suggest that the classical limit is obtained when probability distributions are smooth on scales close to the uncertainty limit of quantum metrology given by $\delta \phi \geq 1/(2 \Delta A)$. 

\section{Conclusions}
\label{sec:concl}

The results in this paper show that the complex conditional probabilities obtained in weak measurements can be interpreted in terms of the dynamics required to optimize the overlap of initial and final state for a range of intermediate states. Weak conditional probabilities oscillating between negative and positive values therefore indicate a difference in the physical properties of initial and final states corresponding to a classical phase space distance. The complex phase itself represents the action of the unitary transformation that maximizes the contribution to the overlap between initial and final state at the intermediate state. 

Effectively, weak conditional probabilities replace the strict logic of classical determinism with an equally strict logic of statistical correlations between measurements that cannot be performed at the same time. In this modified statistical theory, the complex phase represents the dynamic action that would be necessary to overcome the contradictions between the three statements encoded in the initial, intermediate and final states. In terms of quantum information, the complex phase of weak conditional probabilities thus provides a quantitative measure of logical tension between three quantum statements. Since logical tensions greater than $\pi/2$ appear as negative conditional probabilities in the correlations between the results of separate measurements, the violation of classical inequalities in quantum paradoxes can then be understood as a consequence of the high logical tensions characterizing the extreme quantum limit. On the other hand, the classical limit can be defined as the limit of low logical tension, where classical noise reduces the imaginary parts of weak conditional probabilities to the point where they are always much smaller than the (positive) real parts. 

By extending the concept of conditional probabilities to include complex phases identified with the logical tension, weak measurement statistics establish a relation between the non-classical correlations observed in quantum paradoxes and the physics of continuous transformations. This fundamental relation between dynamics and statistic might be the key to a better intuitive understanding of quantum phenomena and their applications.

\section*{Acknowledgment}
Part of this work has been supported by the Grant-in-Aid program of the Japanese Society for the Promotion of Science, JSPS.

\end{document}